\documentclass[12pt]{article}

\usepackage[utf8]{inputenc}
\usepackage{graphicx}
\usepackage{float}
\usepackage{amsmath,amsfonts,amssymb}
\usepackage{esint}
\usepackage{nicefrac}
\usepackage[font=small]{caption}
\usepackage{subcaption}
\usepackage{wasysym}
\usepackage[colorlinks=true, allcolors=blue]{hyperref}
\usepackage{changepage}
\usepackage[T1]{fontenc}
\usepackage{authblk}
\usepackage{siunitx}
\usepackage[square, sort, numbers]{natbib}
\usepackage{commath}
\usepackage[dvipsnames]{xcolor}
\usepackage{tabularx}
\usepackage{booktabs}
\usepackage{wrapfig}
\usepackage{bbm}
\usepackage[overload]{empheq}
\usepackage{lipsum}
\usepackage{color}
\usepackage{soul}
\usepackage{multirow}
\usepackage{ulem}

\title{\vspace{-30mm}\bf\large Measurement and simulation of mechanical and optical properties of sputtered amorphous SiC coatings}

\author[a]{G. Favaro}

\author[b]{A. Amato}
\author[c,d]{F. Arciprete}
\author[a]{M. Bazzan}
\author[d]{E. Cesarini}
\author[c,d]{F. De Matteis}
\author[c,d]{T. H. Dao}

\author[e]{M. Granata}

\author[f]{C. Honrado-Ben\'{i}tez}
\author[f]{N. Guti\'{e}rrez-Luna}
\author[f]{J. I. Larruquert}
\author[a]{G. Lorenzin}
\author[c,d]{D. Lumaca}
\author[a]{G. Maggioni}
\author[g]{M. Magnozzi}
\author[h]{E. Placidi}
\author[c,d]{P. Prosposito}
\author[i]{F. Puosi}

\affil[a]{Università di Padova, I-35131, Padova, Italy}
\affil[b]{Université de Lyon, Université Claude Bernard Lyon 1, CNRS, Institut Lumière Matière, F-69622 Villeurbanne, France}
\affil[c]{Università di Roma Tor Vergata, I-00133 Roma, Italy}
\affil[d]{INFN, Sezione di Roma Tor Vergata, I-00133 Roma, Italy}

\affil[e]{Laboratoire des Mat\'{e}riaux Avanc\'{e}s - IP2I - CNRS, F-69622 Villeurbanne, France}
\affil[f]{GOLD-IO-CSIC, Serrano 144, S-28006 Madrid, Spain}
\affil[g]{Università di Genova, I-16146 Genova, Italy}
\affil[h]{Università di Roma “La Sapienza”, I-00185 Roma, Italy}
\affil[i]{INFN, Sezione di Pisa, I-56127 Pisa, Italy}

\begin{document}

\maketitle

\hrulefill\par

\begin{abstract}
In this work We report on the extensive characterization of amorphous silicon carbide (a-SiC) coatings prepared by physical deposition methods. 
We compare the results obtained on two different sputtering systems (a standard RF magnetron sputtering and a ion-beam assisted sputtering) to seize the impact of two different setups on the repeatably of the results. 

After a  thorough  characterization of structural, morphological, and compositional characteristics of the prepared samples, we focus on a detailed study of the optical and mechanical losses in those materials.
Mechanical losses are further investigated from a microscopic point of view by comparing our experimental results with molecular dynamic simulations of the amorphous SiC structure: first we define a protocol to generate a numerical model of the amorphous film, capturing the main features of the real system; then we simulate its dynamical behaviour upon deformation in order to obtain its mechanical response. 
Our results are discussed within the perspective application of a-SiC as an optical material for high-precision optical experiments and in particular in gravitational wave interferometry.
\end{abstract}

\hrulefill\par

\section{Introduction}
Crystalline silicon carbide (SiC) is one of the most mature among the wide band-gap semiconductors ($2.0 \, \mathrm{eV}\lesssim Eg \lesssim 7.0\, \mathrm{eV}$) for high-temperature, high-power, high-frequency, and radiation hardened applications, due to its favorable physical properties and its commercial availability. One of the most distinctive features of SiC is the rich variety of its crystal forms. SiC is part of a family of materials which exhibit a one-dimensional polymorphism called polytypism.
An almost infinite number of SiC polytypes are possible, and approximately 200 polytypes differentiated by the stacking sequence of each tetrahedrally bonded Si-C bilayer have already been discovered \cite{pensl1993electrical}.
Even though individual bond lengths are nearly identical, the crystal symmetry is determined by the stacking periodicity. 
This variety of structural arrangements is mirrored by the large variability of SiC physical properties such as optical bandgap, thermal and electrical conductivity and so on.

More recently, silicon carbide has attracted a renewed interest in its amorphous form (a-SiC) as a material for optical and optoelectronic applications, as well as for more cutting-edge research. In the case of the amorphous structure, the variety of the allowable bond configurations offered by the SiC polytypism together with the possibility of doping control allow to tune the optical and mechanical properties in situations where high refractive index, mechanical stability and low absorption is required.

In particular, a-SiC could have the potential to address very critical issues in high-precision optical experiments such as interferometric gravitational-wave detectors (GWDs) \cite{Saulson17}, opto-mechanical resonators \cite{Aspelmeyer14}, and frequency standards \cite{Matei17}.
Those experiments and in particular GWDs such as Advanced LIGO \cite{aLIGO}, Advanced Virgo \cite{AdVirgo} and KAGRA \cite{KAGRA} require high-reflection (HR) optical coatings made up of Bragg reflectors of alternating layers of low- and high-refractive-index materials, where a-SiC could be considered as a potential candidate for the high-index layer. In fact, despite the superb optical properties of the coatings used in present GWDs \cite{Degallaix19,amato2019optical,Granata20review}, HR coatings still constitute a severe limitation for further sensitivity improvement in GWDs due to their thermal noise issue, i.e. the presence of thermally-driven random structural relaxations distributing the thermal energy of the normal modes of vibration across a wide frequency range, giving rise to Brownian {\it coating thermal noise} (CTN) \cite{Saulson90,Levin98}. The power spectral density of such thermally induced surface fluctuations is determined by the rate of energy dissipation in the coating material, which is measured by the mechanical loss angle $\varphi$, as stated by the fluctuation-dissipation theorem \cite{Callen52}. Thus, in the last two decades, a considerable research effort has been committed to finding an alternative high-index material featuring both low mechanical and optical losses \cite{Granata20review,Anghinolfi2013,amato2020observation,amato2021optical}.

In this paper we report on the optical and mechanical properties of sputtered amorphous silicon carbide (a-SiC) thin films, and we discuss their potential application in current and future GWDs. 
The main goal of our research is to determine precisely the properties of a-SiC from the standpoint of GWDs requirement, taking into account that the deposition method may strongly affect the structure and thus the properties of the optical film. In order to elucidate how strongly the obtained films are influenced by the deposition methods, we compared coatings produced in two different sputtering systems: (i) a home-made standard laboratory magnetron sputtering apparatus and (ii) an ion-beam sputtering system, which is the technology of choice to produce the tantalum pentoxide (Ta$_2$O$_5$, also known as {\it tantala}, high index) and silicon dioxide (SiO$_2$, {\it silica}, low index), thin coating layers presently used in the mirrors of GWDs \cite{Degallaix19,Pinard17}. We determined the physico-chemical characteristics of the a-SiC thin film samples by a wide range of techniques, with particular attention to their optical and mechanical properties in order to assess the potential of SiC as a material for GWDs. Finally, in order to obtain a microscopic insight on the material, the experimental results are compared with the results of an atomistic calculation simulating the amorphous structure of SiC and its mechanical properties.

\section{Sample production}
\label{sec:production}
Thin layers of amorphous SiC have been deposited on different substrates for different purposes: (i) silicon wafers ($\varnothing$ 25.4 mm, 5-mm thick) and sapphire substrates ($\varnothing$ 25.4 mm, 5-mm thick) as witness samples to measure the optical properties, stoichiometry and structure of the coatings, (ii) amorphous carbon (a-C) with a thin layer of MS gold (Au) to perform accurate Rutherford back-scattering spectrometry experiments, (iii) fused-silica disks ($\varnothing$ 50 mm, 0.1 to 1-mm thick) to measure the coating mechanical properties.

Due to the nature of the mechanical loss measurements, which are particularly sensitive to the experimental conditions, the fused-silica disks were prepared according to the following specific protocol: (i) in order to release the internal stress due to manufacturing and induce relaxation, the disks were annealed in air at 900 $^\circ$C for 10 hours prior to coating deposition; (ii) in order to cancel out the coating-induced curvature that would affect the measurement of their resonant frequencies, thin film of equal thickness have been grown on both their optically polished surfaces.

In some selected cases, the coated samples were annealed to relax internal stresses and improve the coating optical properties. The effects of this treatment from the standpoint of the optical properties will be discussed below.

In order to seize how much the properties of the deposited materials are dependent upon the specific growth conditions, we compared the results obtained using two different sputtering systems. The first one is a standard radio-frequency magnetron sputtering (MS) setup located inside Laboratori Nazionali di Legnaro\footnote{www.lnl.infn.it} (INFN-LNL). The second is an ion-beam sputtering system (IBS) located at the GOLD facility of the Instituto de \'{O}ptica Daza de Vald\'{e}s\footnote{www.io.csic.es} (IO-CSIC).

\subsection{Magnetron sputtering (MS) samples}
In this apparatus a commercial 2-inches SiC target with a purity of 99.999\% was used within a radio-frequency source (Torus® 2C Kurt J. Lesker) operated at 100 W by means of an appropriate RF power supply (RFX600 Advanced Energy). All the processes were performed in an Ar atmosphere. The value of the pressure inside the chamber was around $8 \times 10^{-7}$ mbar before the injection of argon and around $6 \times 10^{-3} $ mbar during the sputtering process. The distance between the target and the substrate was of 14 cm and the flux of argon in the chamber was 17 sccm for all depositions. The substrates are mounted on a rotating sample holder for improved coating thickness homogeneity. The velocity of rotation was set at 30 RPM for all the depositions.
The thickness of the samples deposited in this system ranges from 40 nm to 400 nm, with a deposition rate of 0.017 nm/s.
In order to modify the film stoichiometry, a varying number of Si pieces were placed on the SiC target for some deposition runs. This allowed us to produce a set of samples with different Si/C ratios, in order to explore the influence of the composition on the optical properties.

\subsection{Ion beam sputtering (IBS) samples}
Prior to deposition, the residual pressure inside the coater vacuum chamber was less than $3 \times 10^{-8}$ mbar. Ar was used as sputtering gas, the total pressure was $1 \times 10^{-3}$ mbar during the coating process. Sputtering ions were produced by means of a 3-cm diameter hollow cathode ion gun working with a hollow cathode neutralizer, both without filament in order to minimize contamination. Energy and discharge beam current of the ions were 1.0 keV and 50 mA, respectively, yielding a deposition rate of 0.05 nm/s, as measured with a quartz crystal monitor. A 96.5-mm diameter SiC target of 99.9995\% purity was used, mounted in a multi-target rotating holder cooled down with water. Substrates during rotation were kept at a distance of 21 cm above the target, with the ion beam impinging on the latter at 45 deg. 
Each thin film was grown separately, but all coating samples were grown under identical conditions.

\section{Physico-chemical Characterizations}
Grazing-incidence X-Ray Diffraction (GIXRD) and X-Ray Reflectivity (XRR) analyses have been performed using a Philips MRD diffractometer equipped with a Cu tube operated at 40 kV and 40 mA. The beam was collimated and partially monochromatized by a parabolic multilayer mirror, while the detector was equipped with a parallel plate collimator (PPC) to define the angular acceptance. To perform XRR measurements an additional high-resolution four-bounce Ge (2 2 0) symmetric Bartels monochromator was mounted on the primary optics and the PPC was replaced with a three-bounce symmetric (2 2 0) Ge analyzer.

The morphology of samples were studied by a Veeco Multiprobe Nanoscope IIIa atomic force microscope (AFM). Topographies were acquired in tapping mode (amplitude modulation) by means of a silicon tip with a stiffness of 42 N/m (resonance frequency around 300kHz) and radius of curvature below 7 nm. The measurements were performed in the MBE laboratory in the Physics Department of University of Rome Tor Vergata.

The composition of the films was investigated by Rutherford back-scattering spectrometry (RBS) using a 2.0 MeV 4He+ beam at the Van der Graaf accelerator at the INFN-LNL with a 160-deg scattering angle, normal incidence, and a silicon solid state detector.

\subsection{Results}
 
\subsubsection{X-ray analysis}
\label{SECT_XRR_res}
\begin{figure}
    \centering
    \includegraphics[width=\textwidth]{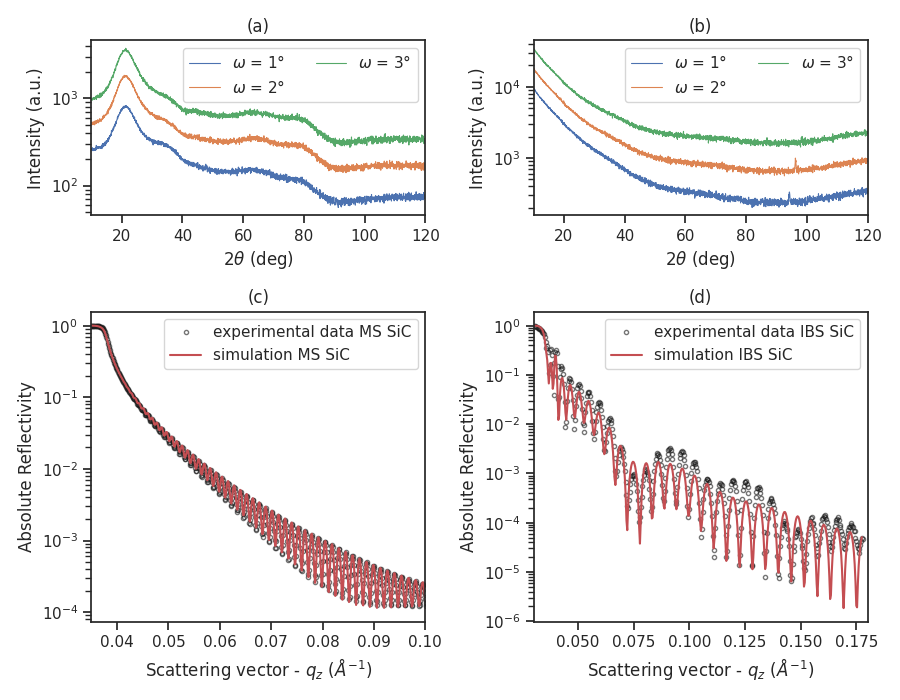}
    \caption{GIXRD and XRR spectra for two a-SiC thin film samples prepared by MS and IBS on different substrates. (a) GIXRD of MS a-SiC on a $\mathrm{SiO_{2}}$ substrate; (b) GIXRD of IBS a-SiC on a $\mathrm{Al_{2}O_{3}}$ substrate; (c) XRR of the same sample as in (a); XRR of the same sample as in (b). The red lines in (c) and (d) are model fits.}
    \label{fig:XRD_XRR}
\end{figure}

In figure \ref{fig:XRD_XRR} are presented the results of the GIXRD and XRR analyses on two representative samples prepared by the MS and IBS deposition systems on two different substrates. The GIXRD spectra were recorded for different values of the incidence angle $\omega = 1,2,3$ deg in order to compare a different penetration depth of the X-ray beam. In all cases the samples did not demonstrate any diffraction peak confirming that the grown SiC layers were  amorphous. Note that the difference between the two spectra reported in Fig \ref{fig:XRD_XRR} (a) and (b) is due to the different substrates: the former is obtained from a film deposited on amorphous silica, which contributes to the diffracted signal with the smooth diffraction pattern typical of this material; the second sample has a crystalline sapphire substrate which does not induce any significant background.

The XRR analysis allows to determine the average density and the surface roughness for the two films. For films thinner than 500 nm, our reciprocal space resolution also allowed us to measure thickness fringes and hence determine the film thickness from an analysis of the curve. Experimental data of the MS sample, shown in Figure \ref{fig:XRD_XRR} (c), is well fit with a simple single-layer model using the REFLEX software \cite{vignaud2019reflex}. The obtained average density, $\rho=(2.82 \pm 0.05)$ g/cm$^3$, is in agreement with previous works \cite{heera1997density}, and the roughness resulted in $\sigma=(2.3 \pm 0.1)$ nm. In the case of the IBS sample, the model is slightly more complex as it required the introduction of a thin low-density cap layer of about 13 nm to model the low-frequency modulations visible in Figure \ref{fig:XRD_XRR} (d). The results for that layer are $\rho=(2.87 \pm 0.01)$ g/cm$^3$ and $\sigma=(0.5 \pm 0.1)$ nm, so the MS and IBS samples have a very similar density but the IBS sample has a much smoother surface. Note however that, besides being deposited on different substrates, those values were obtained in two samples with significantly different thicknesses, $t=(423.1 \pm 0.1)$ nm for the MS sample and $t=(104.3 \pm 0.1)$ nm for the IBS sample, as measured by XRR curves. Those aspects may also contribute to the observed difference in the surface roughness \cite{barabasi1995fractal}.

\subsubsection{Atomic Force Microscopy}
\begin{figure}[h]
	\centering
	\includegraphics[width=\textwidth]{{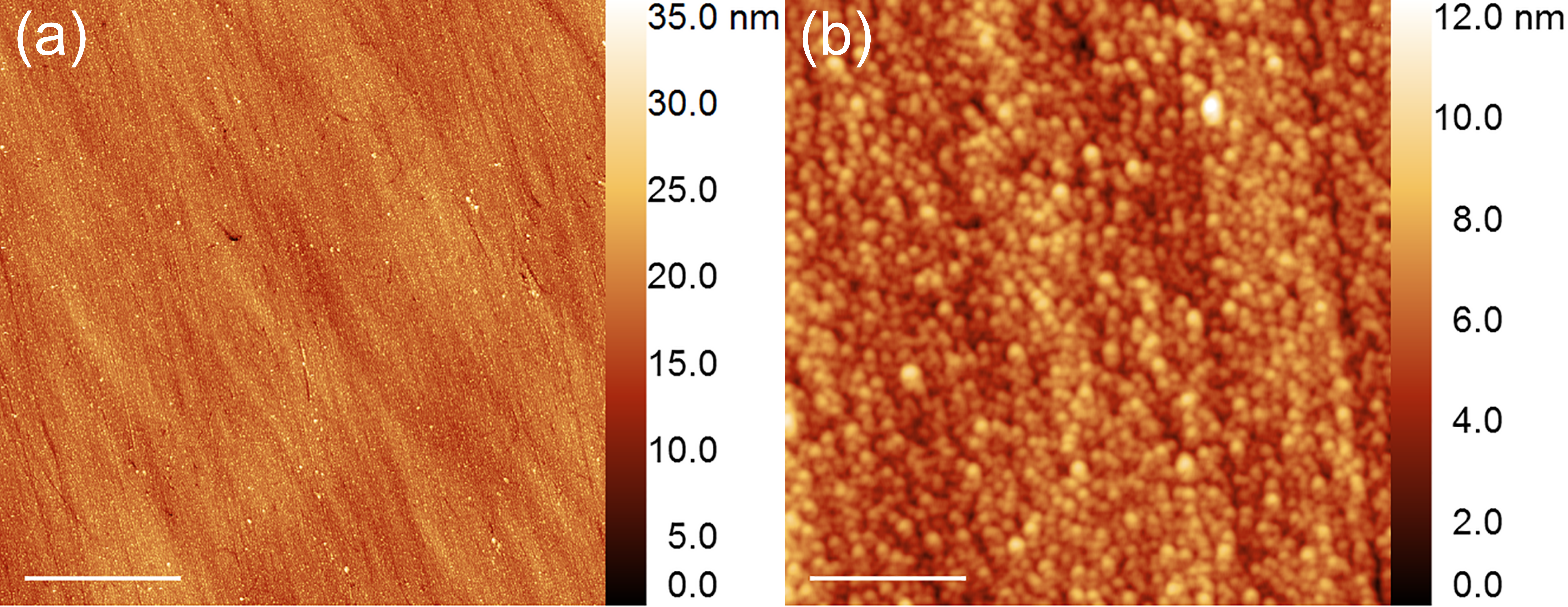}}
	\caption{AFM topographies of SiC surface. The image scale-bars are respectively (a) $\mathrm{5\mu m}$ and (b) 500 nm. }
	\label{Fig_AFM}
	\end{figure}

 MS samples were also characterized by AFM, in order to investigate the details of the surface topography. The results presented in Fig. \ref{Fig_AFM} show that MS samples exhibit a surface morphology characterized by raw textures, probably related to the substrate polishing. In excellent agreement with XRR analysis, the RMS roughness measured by AFM is around 2.2 nm for all the examined maps. On the very surface, small grains with average height and diameter respectively of 5 nm and 20 nm can be observed (as pointed out in panel (b) of Fig. \ref{Fig_AFM}).

\subsubsection{Rutherford Backscattering Spectrometry}
\label{SECT_RBS_res}
\begin{figure}
    \centering
    \includegraphics[width=\textwidth]{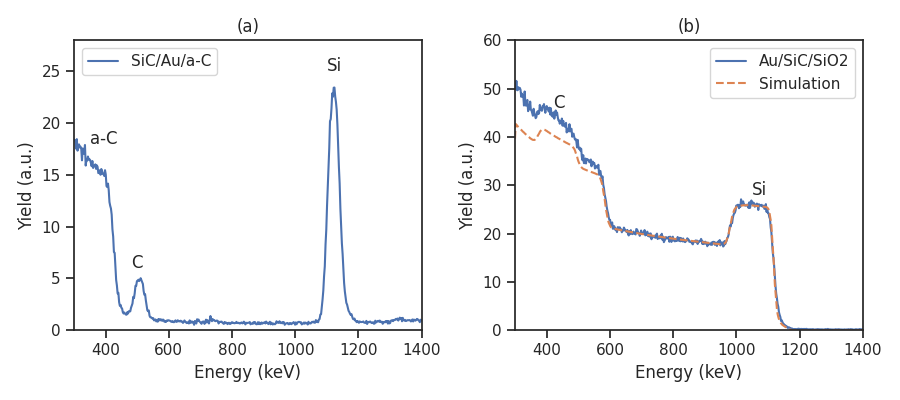}
    \caption{(a) RBS spectrum for MS sample deposited on a Au/C substrate; (b) RBS spectrum for an IBS sample deposited on a $\mathrm{SiO_{2}}$ substrate.}
    \label{fig:RBS}
\end{figure}

Figure \ref{fig:RBS} compares experimental RBS spectra of two a-SiC samples, grown by MS and by IBS on different substrates.

The 50 nm thick MS sample was deposited on a gold-coated a-C substrate; the additional 80 nm thick Au layer allowed us to separate the signal of the a-SiC film from that of the substrate. On Figure \ref{fig:RBS} (a), the first part of the spectrum between 200 keV and 400 keV results from collisions of incident $\alpha$ particles reaching the carbon substrate after losing some energy by passing through the first layers of the sample. The peaks due to collisions with the C and Si atoms in the film are clearly visible (the peak related to collisions with Au atoms is outside the range displayed in the figure). The areal density and the composition of the deposited films is thus obtained by direct integration of the peak area. The average areal density of the measured MS samples is $(2.8 \pm 0.1)\times 10^{17}$ SiC molecules per cm$^2$ which, taking into account the film thickness, results in a density of $\rho = (2.9 \pm 0.1)$ g/cm$^3$, in agreement with the results obtained from the XRR data. Some contamination of other elements has also been detected as traces, mainly O (around 2 mol. percent) while other contamination (Ar from the sputtering atmosphere and Fe from the target holder) are below 1 mol percent. 
The Si/C ratio of an as-deposited MS sample is around 0.74, showing a tendency towards a Si deficiency with respect to the stoichiometric composition. By adding some Si pieces on the target, the Si/C ratio of the deposited films could be varied between 0.74 and 1.15, providing a set of samples with different stoichiometries to be used in the next analysis. 

When different substrates are used, the RBS spectrum cannot be analyzed by direct integration as the signals from the different elements of the film overlap with the substrate signal. Moreover, for non conducting substrate like SiO$_2$, to avoid the sample charging and producing artifacts in the RBS spectrum, an additional Au layer is deposited on top of the sample. This, for example, can be seen in Figure \ref{fig:RBS} (b), which presents the spectrum of an IBS sample grown on a SiO$_2$ substrate. In this case the analysis can be performed by fitting the experimental spectrum with numerical simulations obtained by the RUMP software \cite{DOOLITTLE1985344}, at the price of a higher uncertainty in the elemental analysis. 
In figure \ref{fig:RBS} (b) the signal of Si and C coming from the film is clearly visible, while the incorporation of other elements such as oxygen is not visible. The areal density of the IBS film given by the analysis is $(9.3 \pm 0.1)\times 10^{17}$ SiC molecules per cm$^2$. By dividing by the film thickness, one gets a film density of $\rho = (2.8 \pm 0.2)$ g/cm$^3$, once again in agreement with XRR data. 
Differently from what observed with MS samples, as-deposited IBS samples appear to be more stoichiometric, with a Si/C ratio close to 1. 

\section{Optical Properties}
Optical ellipsometry is an established technique for the study of the optical properties of thin films and in particular of gravitational-wave detectors coatings \cite{Granata20,prato2011gravitational,amato2019optical}.
The optical characterization of a-SiC coating samples was performed using a J. A. Woollam Co. Variable Angle Spectroscopic Ellipsometer (VASE), which exploits the rotating analyzer configuration. The setup is equipped with a computer-controlled compensator (AutoRetarderTM) which allows measurement of $\Psi$ and $\Delta$, both functions of the complex refractive index and of the sample thickness, in the whole range 0-360$^{\circ}$ without ambiguity. 
SE measurements were performed between 300-1700 nm at incidence angles of $55^\circ$, $60^\circ$ and $65^\circ$ in the probed spectral range in order to maximize the sensitivity around the Brewster's angle.
Furthermore, the same setup adjusted in transmission mode allowed for a direct absorbance measurement on the same spectral range. 
Since the absorbance is defined as $\alpha =\frac{4\pi}{\lambda}\kappa$ (where $\kappa$ is the imaginary part of the refractive index), the absorption and SE measurements can be combined in a global minimization analysis using the J. A. Woollam Co. WVASE software package. A Tauc-Lorentz dielectric function ($\varepsilon^{(TL)}$) with an additional Lorentzian component ($\varepsilon^{(L)}$) has been used to model the imaginary part of the dielectric constant of all the samples in the investigated spectral range:

\begin{equation}
\label{EQ_SEmodel}
\begin{split}
\varepsilon_2 (E) & = \varepsilon_2^{(TL)} (E)+\varepsilon_2^{(L)} (E) \\
\varepsilon_2^{(TL)} (E) & = A^{(TL)}\frac{(E-E_g)^2}{E^2}\frac{B_rE_nE}{(E-E_n)^2+(B_rE)^2} \\
\varepsilon_2^{(L)} (E) & = A^{(L)} \frac{CE_0E}{(E-E_0)^2+(CE)^2}
\end{split}
\end{equation} 

The real part of the dielectric constant is then obtained by Kramers-Kronig trasformation. 
The surface roughness is modeled within the effective medium approximation (EMA) as a 50/50 mix of SiC and voids; this layer include also the effect of surface oxide \cite{Zollner}.

\subsection{Results}
\label{SECT_opt_res}

\begin{figure}[h]
	\centering
	\includegraphics[width=8cm]{{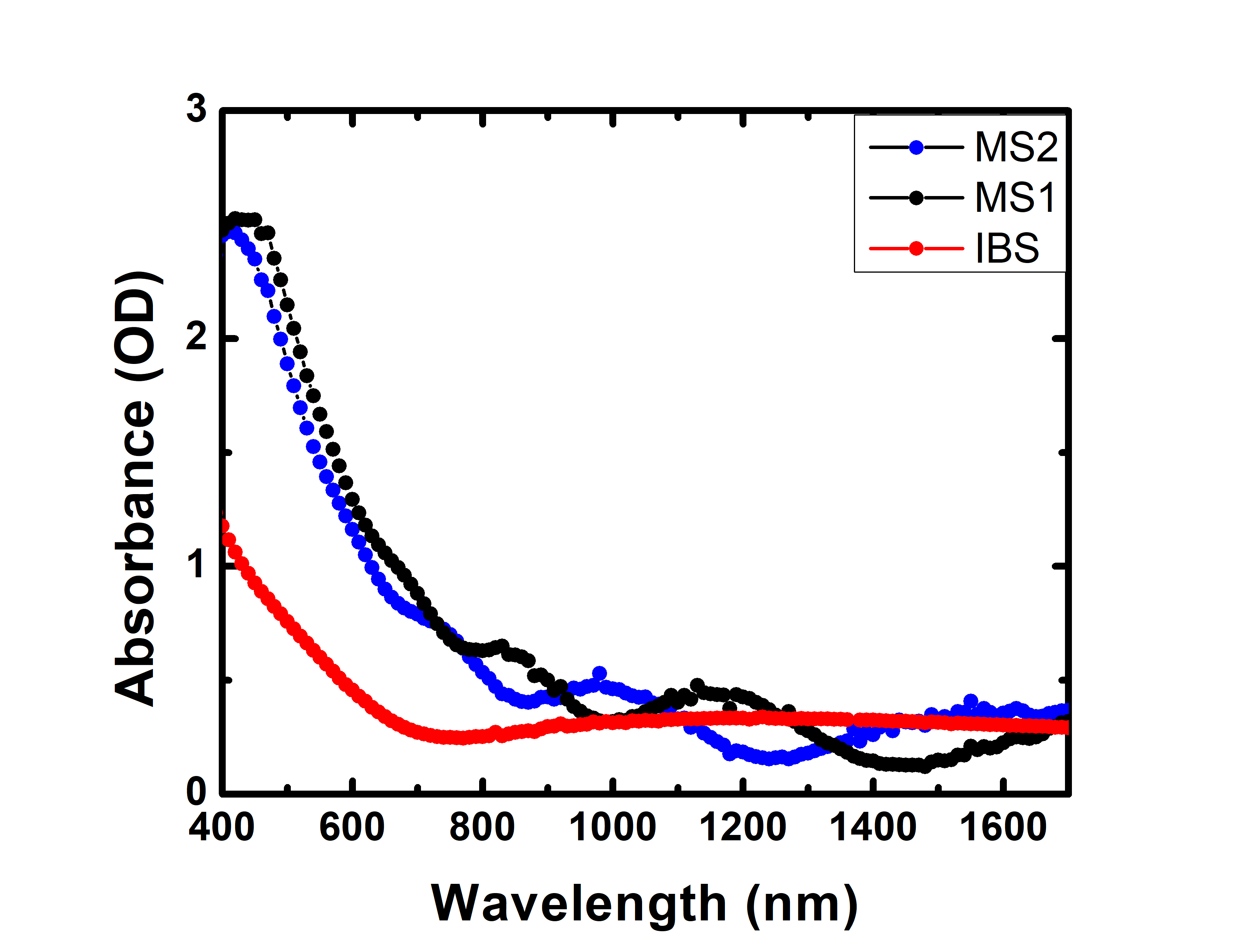}}
	\caption{Absorption spectra of two a-SiC samples grown via MS on a 0.1 mm thick silica substrate and of an a-SiC sample grown via IBS on a 5 mm thick sapphire substrate. }
	\label{AbsEner}
\end{figure}

\begin{figure}
	\centering
	\includegraphics[width=8cm]{{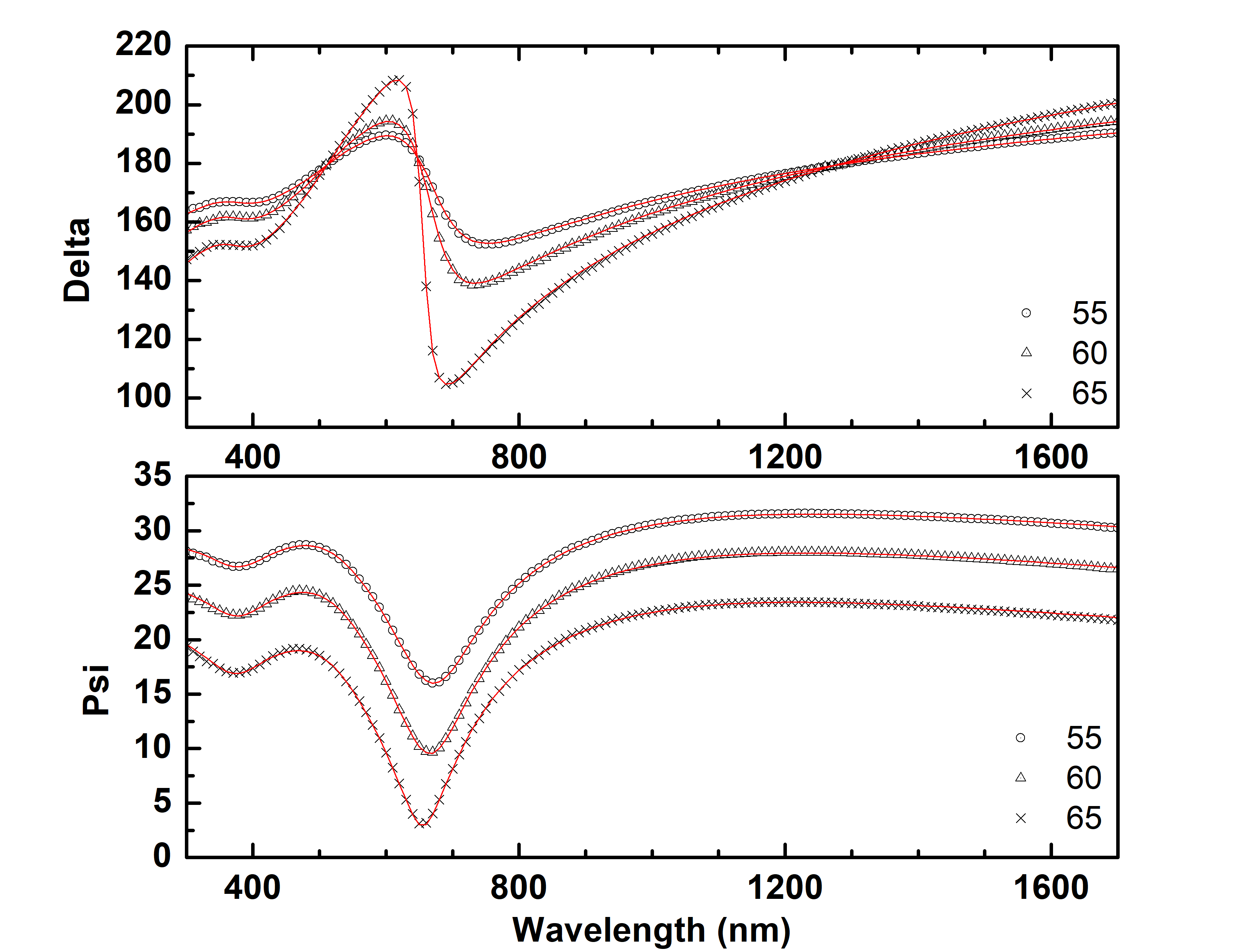}}
	\caption{Spectroscopic ellipsometry of an a-SiC sample grown by means of IBS on a 5 mm thick sapphire substrate. The experimental data (symbols) was measured at an incidence angle of 55$^\circ$, 60$^\circ$ and 65$^\circ$. The red lines are fit curves obtained by the global optimization of the dielectric model function.}
	\label{Se_IBS}
\end{figure}

\begin{table}
    \begin{center}
    \caption{Fitting parameters for the Tauc-Lorentz (TL) plus Lorentzian (L) dielectric model for the a-SiC samples grown via MS and IBS.}
\begin{tabular}{|c|c|c|c|c|}
\hline
\multicolumn{2}{|c|}{Sample} &\verb|MS 1| &\verb|MS 2|&	\verb|IBS| \\\hline
\multicolumn{2}{|c|}{Substrate}  & SiO$_2$ 0.1 mm &	SiO$_2$ 0.1  mm & Al$_2$O$_3$ 5 mm \\\hline\hline
\multicolumn{2}{|c|}{MSE}&	7.7	& 8.3	& 1.4	\\\hline\hline
\multicolumn{2}{|c|}{Roughness(nm) } &	8.3 $\pm$ 0.2 
& 5.8$\pm$0.2 
& 1.50$\pm$0.03
\\\hline	
\multicolumn{2}{|c|}{Thickness (nm) }& 401.8 $\pm$ 0.7 
& 478.8 $\pm$ 0.8 
& 101.48 $\pm$ 0.03 
\\\hline\hline
\multirow{4}{1em}{TL} & A$^{\mbox{{\tiny(TL)}}}$ &	150 $\pm$ 9
&	208  $\pm$ 10 
&	244 $\pm$ 5 
\\\cline{2-5}
 & E$_n$ (eV) &	5.7 $\pm$ 0.2
&	5.3 $\pm$ 0.1
&	4.80 $\pm$ 0.04 
\\\cline{2-5}
 & E$_g$ (eV) & 	1.79 $\pm$ 0.04
& 2.01 $\pm$ 0.03
&	2.25 $\pm$ 0.01 
\\\cline{2-5}
& B$_r$ (eV) &	7 $\pm$ 1 
&	9.1 $\pm$ 1.0
&	6.8 $\pm$ 0.2
\\\hline\hline
\multirow{3}{1em}{L}& A$^{\mbox{{\tiny(L)}}}$ &	1.42 $\pm$ 0.09 
&	1.73 $\pm$ 0.07 
&	2.74 $\pm$ 0.04
\\\cline{2-5}
& E$_0$ (eV)& 	2.26 $\pm$ 0.05 
& 2.33 $\pm$ 0.03 
&	2.66 $\pm$ 0.01 
\\\cline{2-5}
& C (eV)&		1.40 $\pm$ 0.03 
& 1.29 $\pm$ 0.03
&	1.44 $\pm$ 0.01 
\\\hline
\end{tabular}
    \label{FitParam}
    \end{center}
\end{table}

Figure \ref{AbsEner} shows the absorption spectra of two MS samples deposited on a 0.1 mm thick silica substrate and of an IBS sample deposited on a 5 mm thick sapphire substrate. The nominal thicknesses of the a-SiC layers were respectively 500 nm, 400 nm and 100 nm. 
A typical comparison  between the experimental SE data and the fitted model function is shown in Figure \ref{Se_IBS} for the IBS sample grown on a sapphire substrate. The parameters of the dielectric model (Eq. \eqref{EQ_SEmodel}) are reported in Table \ref{FitParam} for all samples. 

\begin{figure}[h]
	\centering
	\includegraphics[width=\textwidth]{{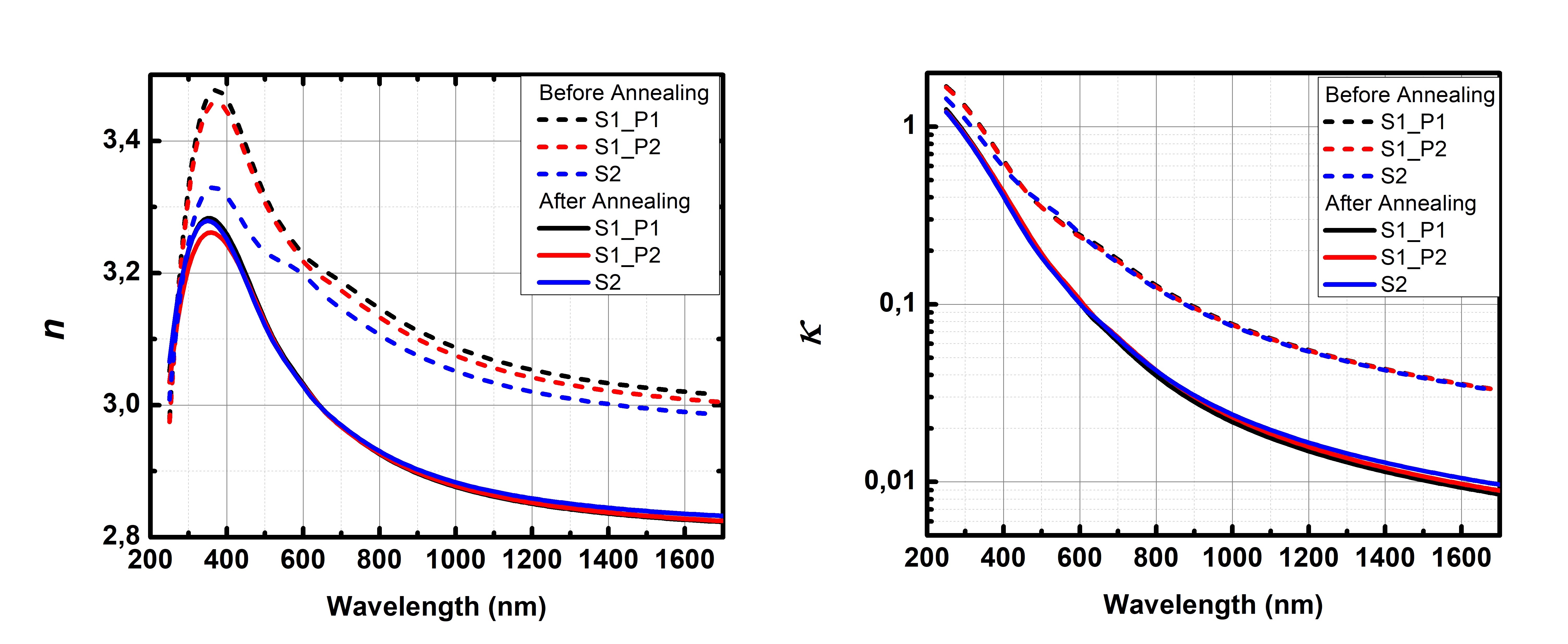}}
	\caption{Real (left) and imaginary (right) part of the refractive index of two MS a-SiC samples. Sample S1, which was measured at two different positions P1 (black) and P2 (red), and sample S2 (blue) were annealed for 10 h at 500 $^\circ$C with heating and cooling rates of 100 $^\circ$C/h: the optical constants are reported before (dashed) and after (full) annealing. }
	\label{Annealing}
\end{figure}

Optical data indicate that the  optical gap Eg ranges from 1.8 to 2.25 eV, and confirm the higher roughness of MS samples as compared to that of the IBS samples. 
Exemplary plots of the refractive index and absorption coefficient as a function of the wavelength  measured in two different positions and on two different MS samples are reported in Fig. \ref{Annealing}.
Although a certain variability is present from point to point and on the two different samples, the values of the  $n$ and $\kappa$ optical constants at 1064 nm are around 3.0 and 0.1, respectively. Those  results fall within the quite wide range of optical-constant data reported in literature for SiC, corresponding to different deposition techniques and conditions \cite{Larruquert}.  Yet this result indicates that our as-deposited SiC samples are characterized by a very high absorption level, several orders of magnitude higher than typical values encountered in other optical coatings for such as e.g. Tantala. 

Other MS samples with different compositions Si/C ranging from  0.74 to 1 were analyzed as well (data not shown), not providing a clear trend with respect to composition. 
We conclude that the optical properties of the different as-deposited samples exhibit a certain degree of variability up to 10\%, which can not be unambiguously correlated to the composition and deposition technique.

After 6 month, the two MS samples have been annealed at 500 $^\circ$C in air for 10 h  with a ramp rate of 1.6 $^\circ$C/min. The temperature was chosen in order to allow relaxation of the amorphous phase, avoiding any micro-crystallization process \cite{Musumeci97}. Figure \ref{Annealing} shows the effect of the 
annealing on the optical constants for the MS sample 1 (measured in two points) and MS sample 2. Both the real and imaginary part of the refractive index decrease to about 2.9 and 0.02 respectively and become much less scattered, hinting that the different optical behaviors observed in the as-deposited samples are related to an initial unrelaxed state bearing a certain level of spatial inhomogeneity. 
Still, in spite of a considerable decrease of the imaginary part of the refractive index $\kappa$ upon annealing, the absorption remain way above the level of $10^{-6} - 10^{-7}$ required for GW detectors.

\section{Mechanical properties}
Three fused-silica disks with a nominal diameter of 50 mm were used for the characterization of the coating mechanical properties: disks A and B, 1-mm thick, coated with IBS a-SiC thin films, were measured at LMA; disk C, 0.35-mm thick, coated with MS a-SiC thin films, was measured at Universit\`{a} degli Studi di Roma Tor Vergata (UniToV). All the samples had nearly-stoichiometric composition, as verified via RBS measurements (see Sect. \ref{SECT_RBS_res}). XRR data was used to estimate the density and thickness of the thin films (see Sect. \ref{SECT_XRR_res}). 

\begin{figure}
\centering
	\includegraphics[width=8cm]{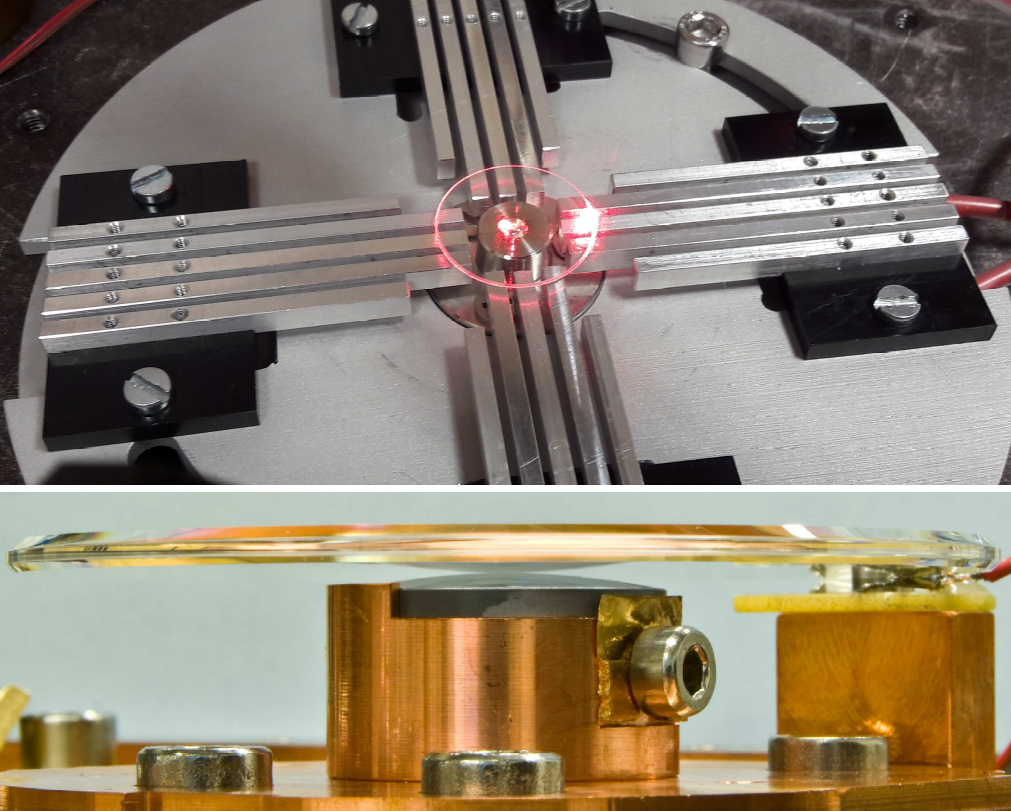}
	\caption{GeNS systems used at Universit\`{a} degli Studi di Roma Tor Vergata (top) and at Laboratoire des Mat\'{e}riaux Avanc\'{e}s (bottom) to measure the mechanical properties of thin films.}
	\label{FIG_GeNSs}
\end{figure}

We used the ring-down method \cite{Nowick72} to measure the frequency $f$ and ring-down time $\tau$ of the first vibrational modes of each disk, before and after the coating deposition, and calculated the coating loss angle
\begin{equation}
\label{EQ_coatLoss}
\varphi_c = \frac{\varphi + (D-1)\varphi_0}{D} \ ,
\end{equation}
where $\varphi_0 = (\pi f_0 \tau_0)^{-1}$ is the measured loss angle of the bare substrate, $\varphi = (\pi f \tau)^{-1}$ is the measured loss angle of the coated disk. $D$ is the measured frequency-dependent \textit{dilution factor} \cite{Li14},
\begin{equation}
\label{EQ_dilFact}
D = 1 -  \frac{m_0}{m} \left( \frac{f_0}{f} \right)^2 \ ,
\end{equation}
where ($m_0$, $m$) is the disk mass with and without the coating, respectively. We measured modes from $\sim$2.5 kHz to $\sim$38.5 kHz for disks A and B and from $\sim$ 900 Hz  to $\sim$18 kHz for disk C, in a frequency band which partially overlaps with the detection band of ground-based gravitational-wave detectors ($10 - 10^4$ Hz). In order to avoid systematic damping from suspension and ambient pressure, we used two clamp-free, in-vacuum Gentle Nodal Suspension (GeNS) systems \cite{Cesarini09}, shown in Fig. \ref{FIG_GeNSs}. This kind of system is currently the reference solution of the Virgo and LIGO Collaborations for performing loss angle measurements on thin films \cite{Granata20,Vajente17}.

The diameter of all disks was measured with a caliper. Mass values $m_0$ and $m$ of disks A and B were measured with an analytical balance, before and after coating deposition. Mass values of disk C were estimated from known properties: substrate and coating thicknesses measured via spectroscopic ellipsometry (see Section \ref{SECT_opt_res}), top and bottom surface areas estimated from diameter measurements, substrate density 2.20 g/cm$^3$ taken from the literature \cite{corning7980}, coating density measured via XRR analysis.

By definition, the measured dilution factor $D$ of Eq. (\ref{EQ_dilFact}) is very sensitive to variations of frequencies and masses \cite{Granata21}, so we took special care to prevent the curvature of the disks induced by the presence of the coating layers by performing a double-deposition on both the substrate faces, as explained in section \ref{sec:production}, and monitored the temperature of the disks during the ring-down measurements. Indeed, the frequency ratio in Eq. (\ref{EQ_dilFact}) depends on the sample Young modulus, which is in turn temperature dependent. To keep track of this issue, each GeNS setup has its temperature probe: right under the copper base plate of the LMA system (visible in the foreground of Fig. \ref{FIG_GeNSs}), and fixed to the optical bench at UniToV. The GeNS system at LMA is installed in a clean room where the temperature is stabilized to (21.9 $\pm$ 0.5) $^\circ$C, whereas the temperature of the GeNS system at UniToV is not stabilized. Therefore, in order to obtain resonant frequency values $f(T_0)$ at a reference temperature $T_0=$ 22.9 $^\circ$C, $f(T)$ values measured at UniToV were corrected according to the following equation,
\begin{equation}
f(T_0) = \frac{f(T)}{1 + \eta T + q} \ ,
\end{equation}
where $\eta \sim 9.5 \times 10^{-3}$ K$^{-1}$ and $q \sim 1.9 \times 10^{-5}$ are coefficients we previously measured on nominally identical fused-silica disks. This correction is critical, whenever mode frequencies are measured in a system where temperature may drift \cite{Granata21}. Further details about our GeNS systems are available elsewhere \cite{Cesarini09,Granata16}.

\subsection{Results}
\label{SECT_Mech_Prop}
\begin{figure}
\centering
	\includegraphics[width=\textwidth]{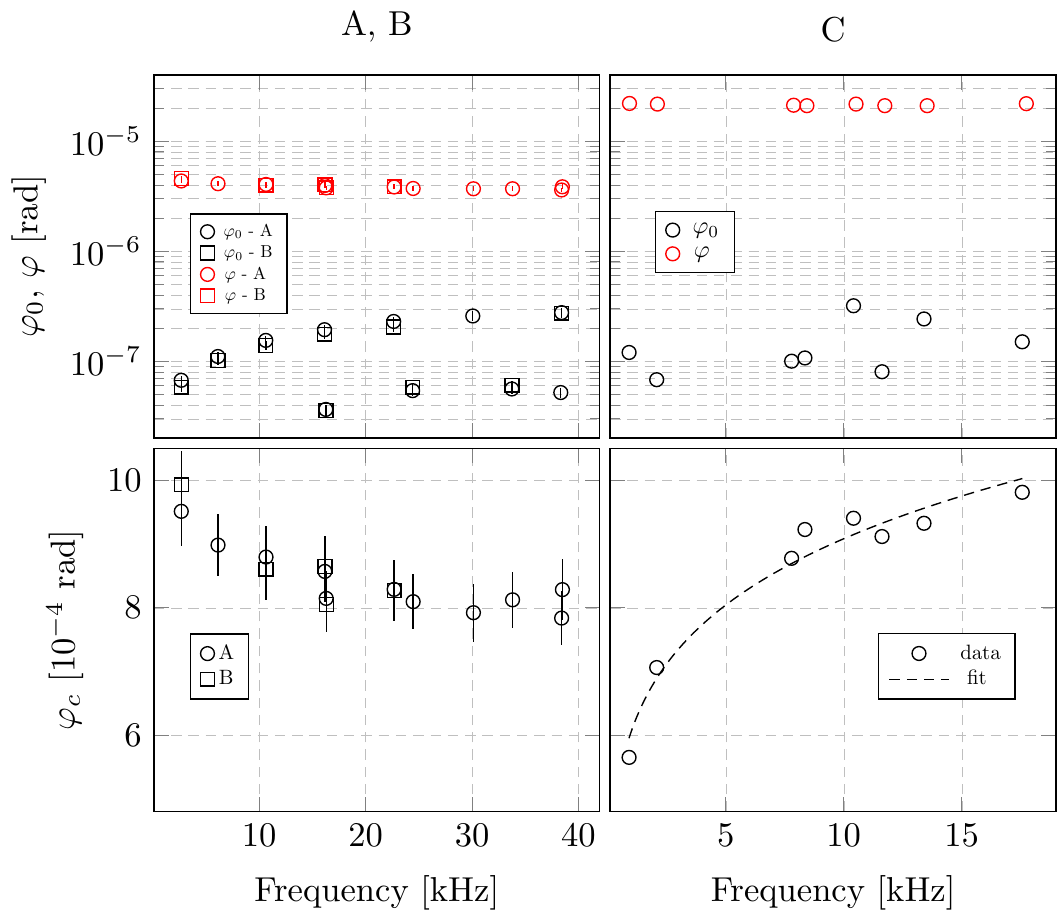}
	\caption{Characterization of mechanical loss of disks A and B, coated with IBS a-SiC thin films ({\it left column}), and C, coated with MS a-SiC thin films ({\it right column}), as a function of frequency. {\it Top row:} measured loss angle before and after deposition of sputtered a-SiC thin films ($\varphi_0$ and $\varphi$, respectively). {\it Bottom row:} coating loss angle $\varphi_c$ of as-deposited IBS ({\it left}) and MS ({\it right}) a-SiC thin films (see Eq.(\ref{EQ_coatLoss}) for more details); for the MS a-SiC coating samples, the best-fit power-law model of Eq.(\ref{EQ_powerLaw}) is also shown (dashed line). }
	\label{FIG_mechLoss}
\end{figure}

The main features of the disks used for the measurements of the coating mechanical properties are presented in Table \ref{TABLE_samples}.
\begin{table}
\caption{\label{TABLE_samples} Disks used to characterize the mechanical properties of the a-SiC sputtered thin films: diameter $\varnothing$, thickness $d_0$, mass $m_0$ before coating, mass $m$ after coating, coating thickness $d_i$ on each side ($i = 1, 2$). Mass values $m_0$ and $m$ for disk C are estimations derived from known properties (see text for more details). Disks A and B were coated with IBS thin films, disk C with MS thin films.}
\centering
\begin{tabular}{ccccc}
    \hline
	& A & B & C \\
	$m_0$ [g] & 4.6453 $\pm$ 0.0003 & 4.6448 $\pm$ 0.0003 & 1.52 $\pm$ 0.18 \\
	$m$ [g] & 4.6472 $\pm$ 0.0002 & 4.6468 $\pm$ 0.0002 & 1.52 $\pm$ 0.18 \\
	$\varnothing$ [mm] & 49.93 $\pm$ 0.01 & 49.92 $\pm$ 0.01 & 50.8 $\pm$ 0.1 \\
	$d_0$ [mm] & 1.08 $\pm$ 0.01 & 1.08 $\pm$ 0.01 & 0.34 $\pm$ 0.04\\
	$d_1$ [nm] & 220 $\pm$ 5 & 221 $\pm$ 5 & 430.7 $\pm$ 0.3\\
	$d_2$ [nm] & 220 $\pm$ 5 & 222 $\pm$ 5 & 422.5 $\pm$ 0.3\\
	\hline
\end{tabular}
\end{table}
Figure \ref{FIG_mechLoss} shows the measured loss angles as a function of frequency: the loss angle of the IBS a-SiC coatings decreases from about 10 to $8 \times 10^{-4}$ rad, whereas the loss angle of the MS a-SiC coatings increases from about 6 to $10 \times 10^{-4}$ rad. Such different behavior might be explained by the nature of the coating samples, due to the different growth techniques.
For instance, AFM measurements (see Fig. \ref{Fig_AFM}) showed that MS samples present a grainy structure which is not observed in IBS ones, indicating a difference in the film microstructure which could possibly reflect the difference in their mechanical behaviour.

For the MS a-SiC samples only, we could fit a power-law model \cite{Gilroy81,Travasso07,Cagnoli18}
\begin{equation}
\label{EQ_powerLaw}
\varphi_c(f) = a f^b
\end{equation}
to our data via least-squares linear regression. Table \ref{TABLE_mechProp} lists the obtained best-fit parameters $(a,b)$, together with the best-fit estimations of coating Young modulus $Y$ and Poisson ratio $\nu$ of the IBS a-SiC coatings, obtained by fitting finite-element simulations to the measured dilution factor via least-squares numerical regression \cite{Granata20, Granata16}.
\begin{table}
\caption{\label{TABLE_mechProp} Measured mechanical properties of a-SiC sputtered thin films: density $\rho$ (see Section \ref{SECT_XRR_res}), best-fit Young modulus $Y$ and Poisson ratio $\nu$ of the IBS samples and best-fit parameters $(a,b)$ of the power-law model used to describe the data of the coating loss angle of MS samples (see Eq.(\ref{EQ_powerLaw}) for more details).}
\begin{tabular}{cccccc}
    \hline
	& $\rho$ [g/cm$^3$] & $Y$ [GPa] & $\nu$ & $a$ [$10^{-4}$ rad] & $b$\\
	disk A & \multirow{2}{*}{2.87 $\pm$ 0.01} & 273 $\pm$ 1 & 0.22 $\pm$ 0.02\\
	disk B & & 276 $\pm$ 2 & 0.20 $\pm$ 0.01\\
	disk C & 2.82 $\pm$ 0.05 & & & 1.8 $\pm$ 0.4 & 0.18 $\pm$ 0.03\\
	\hline
\end{tabular}
\end{table}

Overall, the loss angle of the sputtered a-SiC coatings is a few times higher than that of the as-deposited high-index layers of current GWDs \cite{Granata20}.

\section{Molecular Dynamic simulations}
\subsection{Model and methods}
In order to model from a microscopic point of view the observed mechanical properties of our films, we developed a numerical analysis of the structure and properties of amorphous SiC.
Classical Molecular Dynamics simulations were carried out using LAMMPS software \cite{lammps}.  We model SiC using an effective interatomic potential consisting of two- and three-body interactions. The potential energy of the system is given by: 
\begin{equation}
V=\sum_{i<j}V_{ij}^{(2)}(r_{ij})+\sum_{i,j<k}V_{jik}^{(3)}(r_{ij},r_{ik})
\end{equation}
The two-body term $V_{ij}^{(2)}(r_{ij})$ includes steric repulsion between ions, Coulomb repulsion due to charge-transfer effects between ions, charge-dipole interactions due to electronic polarizability of ions and van der Waals interactions. 
The three-body term $V_{jik}^{(3)}(r_{ij},r_{ik})$ is written as the product of a bond stretching and bending dependence.
Full details about the interatomic potential, including the specific values of the parameters, are given in \cite{SiC_potential}. 

Each simulations consists of 10648 atoms contained in a cubic box with periodic boundary conditions. The structure of an amorphous solid sample was obtained by simulating a steep cooling from a high-temperature liquid. Silicon carbide crystal is first equilibrated at $300\,\mbox{K}$ and then rapidly heated to  $5000\,\mbox{K}$. The melting of the crystal, resulting in a liquid structure, is confirmed after examining the pair distribution function. The liquid at $5000\,\mbox{K}$ is equilibrated for $50\,\mbox{ns}$ and then cooled down to $0\,\mbox{K}$ at constant  rate within the NPT ensemble (constant number of particles $N$, pressure $P$ and temperature $T$). Different values of the cooling rate have been considered. During the quench run, configurations at the temperatures of interest were collected, equilibrated again for $50\,\mbox{ps}$ and finally energy minimized. 
With this protocol, we generated 20 independent samples for each cooling rate. 

Macroscopic elastic moduli are obtained following the quasistatic deformation protocol \cite{Tsamado_PRE09}. The system is deformed in small increments  $\delta \epsilon = 10^{-5}$ followed by energy minimization at fixed applied strain via a conjugate gradient method. The deformation is performed until the applied strain reached $\epsilon=0.005$, which is small enough to ensure that the systems deforms elastically and the corresponding stress-strain curve is linear. The values of the bulk modulus $K$ and shear modulus $G$ are obtained from the slopes of the curves in the case of isotropic bulk deformation and shear deformation respectively. Assuming isotropic behavior, the Young modulus $E$ and the Poisson ratio $\nu$ can be estimated as $E=9KG/(3K+G)$ and $\nu=(3K-2G)/(2(3K+G))$ respectively. 

Mechanical loss is directly obtained using Dynamical Mechanical Spectroscopy (DMS) on the simulated system. DMS is performed by imposing to the simulation box a sinusoidal tensile strain $\epsilon_{ii}(t)=\epsilon_0 \sin (\omega t)$ in the $i$-direction ($i=X,Y,Z$) and measuring the corresponding tensile stress along the same direction, $\sigma_{ii}$. No deformation of the simulation box takes place in the other directions. The results were averaged over all the three directions. We fixed the strain amplitude $\epsilon_0=0.01$, such that the deformation is in the linear elastic regime. The frequency dependent quality factor $Q(f)$ is determined by the ratio   $Q(f)=E'(f)/E''(f)$ where $E'(f)$ and $E''(f)$ are the storage and the dissipative parts of the dynamic elastic modulus. The inverse quality factor $Q^{-1}$ is a direct estimate of the loss angle $\varphi$ since $Q^{-1}\approx \varphi$ holds in the case of small mechanical loss.

\subsection{Results}
\subsubsection{Competition between crystallization and glass formation}
\begin{figure}[t]
\begin{center}
\includegraphics[width=\textwidth]{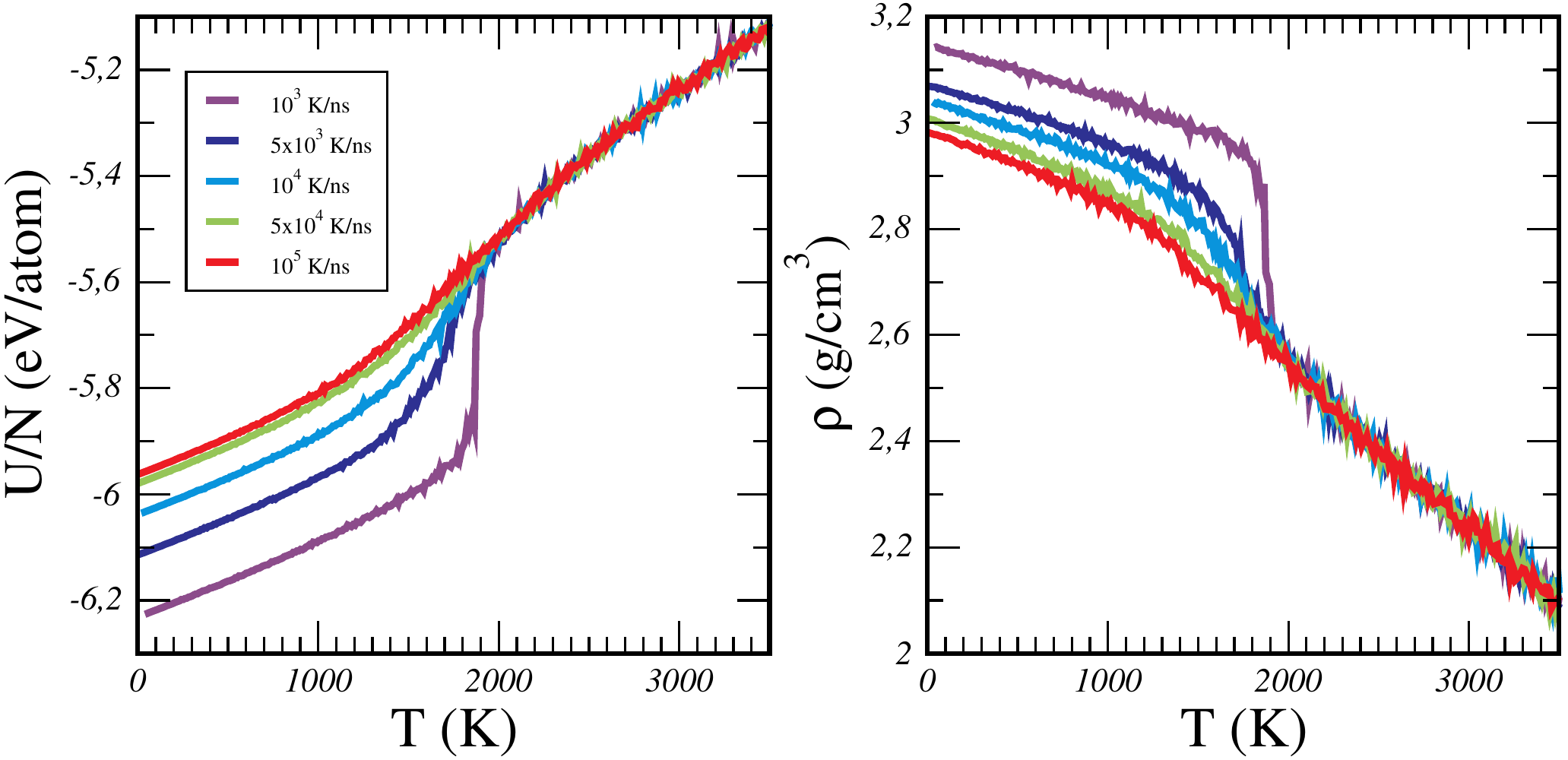}
\end{center}
\caption {The evolution of the potential energy per atom $U/N$ (left ) and density $\rho$ (right panel) upon cooling for  different quench rates, ranging from $10^3\,\mbox{K/ns}$ to $10^5\,\mbox{K/ns}$.  }
\label{figSimCooling}
\end{figure}

In Figure \ref{figSimCooling} we show for a given sample the evolution of the potential energy per atom $U/N$ (left panel) and density $\rho$ (right panel) upon cooling for  different quench rates, ranging from $10^3\,\mbox{K/ns}$ to $10^5\,\mbox{K/ns}$.  We observe that for the slowest cooling rate $10^3\,\mbox{K/ns}$ both the energy and the density  exhibit a sharp change at about $2000\,\mbox{K}$. This is the signature of crystallization occurring in the system, confirmed by direct visual inspection of the sample which features apparent ordered structures (not shown). Crystallization occurs also for a faster rate $5\times10^3\,\mbox{K/ns}$, though changes are less marked. This behavior is common to all the 20 samples we considered. 
As the experimental results indicate that our samples are amorphous, we have an indication that a proper representation of our SiC films must be obtained by employing sufficiently fast cooling rates. 

\begin{figure}
\begin{center}
\includegraphics[width=8cm,trim={3cm 3cm 4cm 2cm},clip]{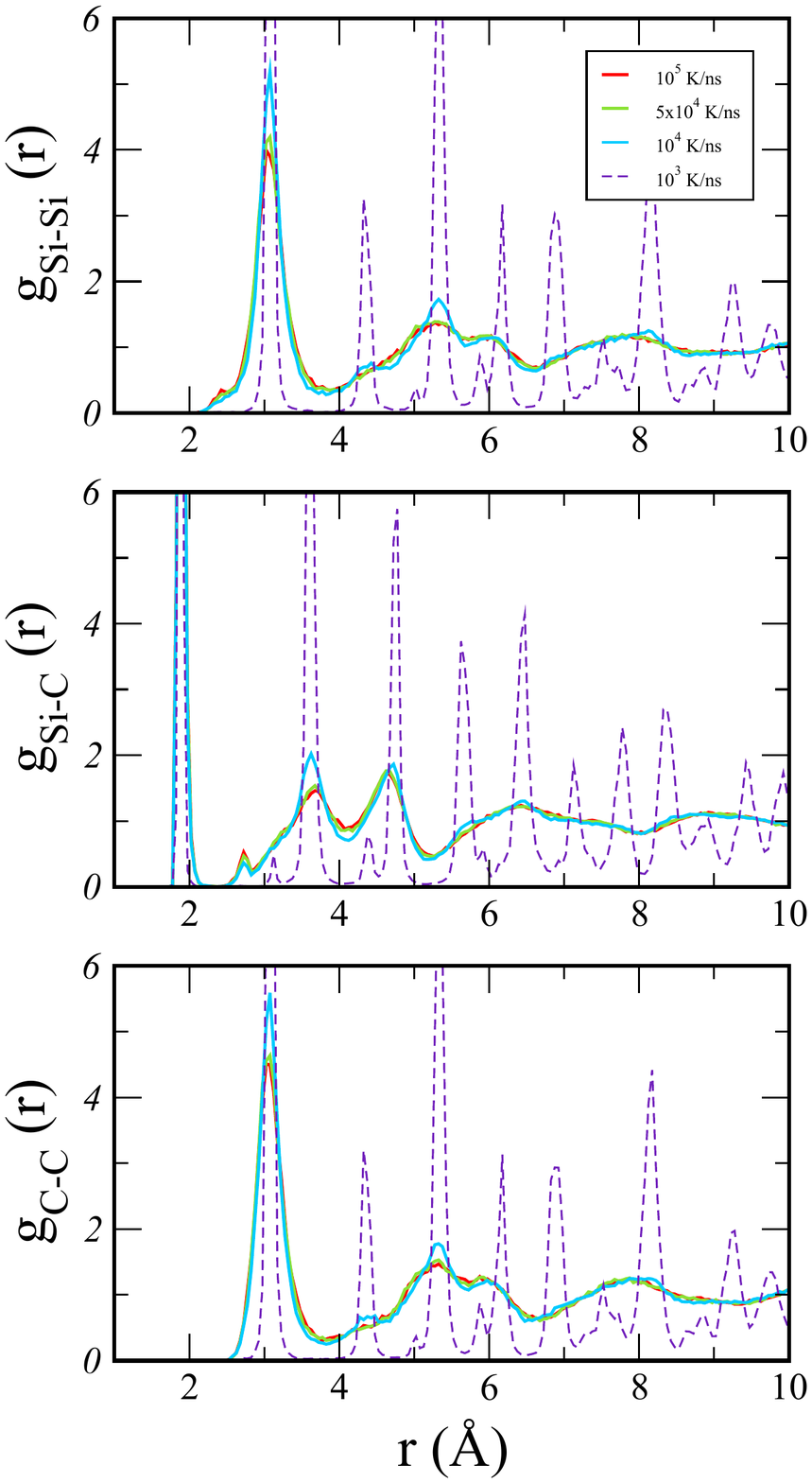}
\end{center}
\caption {Partial pair distribution function at  $T=300\,\mbox{K}$. Data are shown for different quench rates  and corresponds to the average over 20 independent samples for each rate. }
\label{figSimGdR}
\end{figure}

Visual inspection of configurations is not sufficient to understand the evolution for rates faster than $10^5\,\mbox{K/ns}$. Therefore, we resort to statistical structural analysis based on the partial pair distribution function $g_{X-Y}(r)$ with $X,Y=\mbox{Si,C}$ which is plotted in Figure \ref{figSimGdR} for configuration at $T=300\,\mbox{K}$. Data are shown for different quench rates $q=10^5\,\mbox{K/ns},5\times10^4\,\mbox{K/ns}, 10^4\,\mbox{K/ns}$ and corresponds to the average over 20 independent samples for each rate. For comparison we  also show the pair distribution function for a fully crystallized sample produced with a rate $q=10^3\,\mbox{K/ns}$. For the two fastest rates $10^5\,\mbox{K/ns}$ and $5\times10^4\,\mbox{K/ns}$ the pair distribution functions points to an amorphous structure. Conversely, for  $10^4\,\mbox{K/ns}$ the incipient splitting of the peak corresponding to the second coordination shell suggest the development of order in the system which is presumably of crystalline nature. 
A rough estimate of the crystalline fraction in our system can be obtained using the method proposed in \cite{Maras_CPC16}, which allows one to identify cubic and hexagonal crystalline order.  We find that for $10^4\,\mbox{K/ns}$ the crystalline fraction in the 20 independent samples range from $50\%$ to $80\%$ whereas it is less than $10\%$ and less than $5\%$ in the samples produced with rates $5\times10^4\,\mbox{K/ns}$ and  $10^5\,\mbox{K/ns}$ respectively. 

\subsubsection{Mechanical properties}
In this section we report the results on the modelling of the dynamical response of amorphous SiC samples. 
We focus on samples quenched with a cooling rate $5\times10^4\,\mbox{K/ns}$. This choice represents a compromise between a fast quench rate, which limits the crystalline fraction in the system, as previously discussed, and a slow quench rate which is known to results in more relaxed samples, with lower mechanical losses \cite{Puosi_PhysRevRes2019}. Further, SiC samples produced with this cooling rate have a density   $\rho=2.98\,\mbox{g/cm}^3$ which is in reasonable agreement with the experimental value of a-SiC thin films grown via IBS and MS techniques. 

\begin{figure}
\centering
\includegraphics[width=8cm,trim={2cm 12cm 2cm 4cm}]{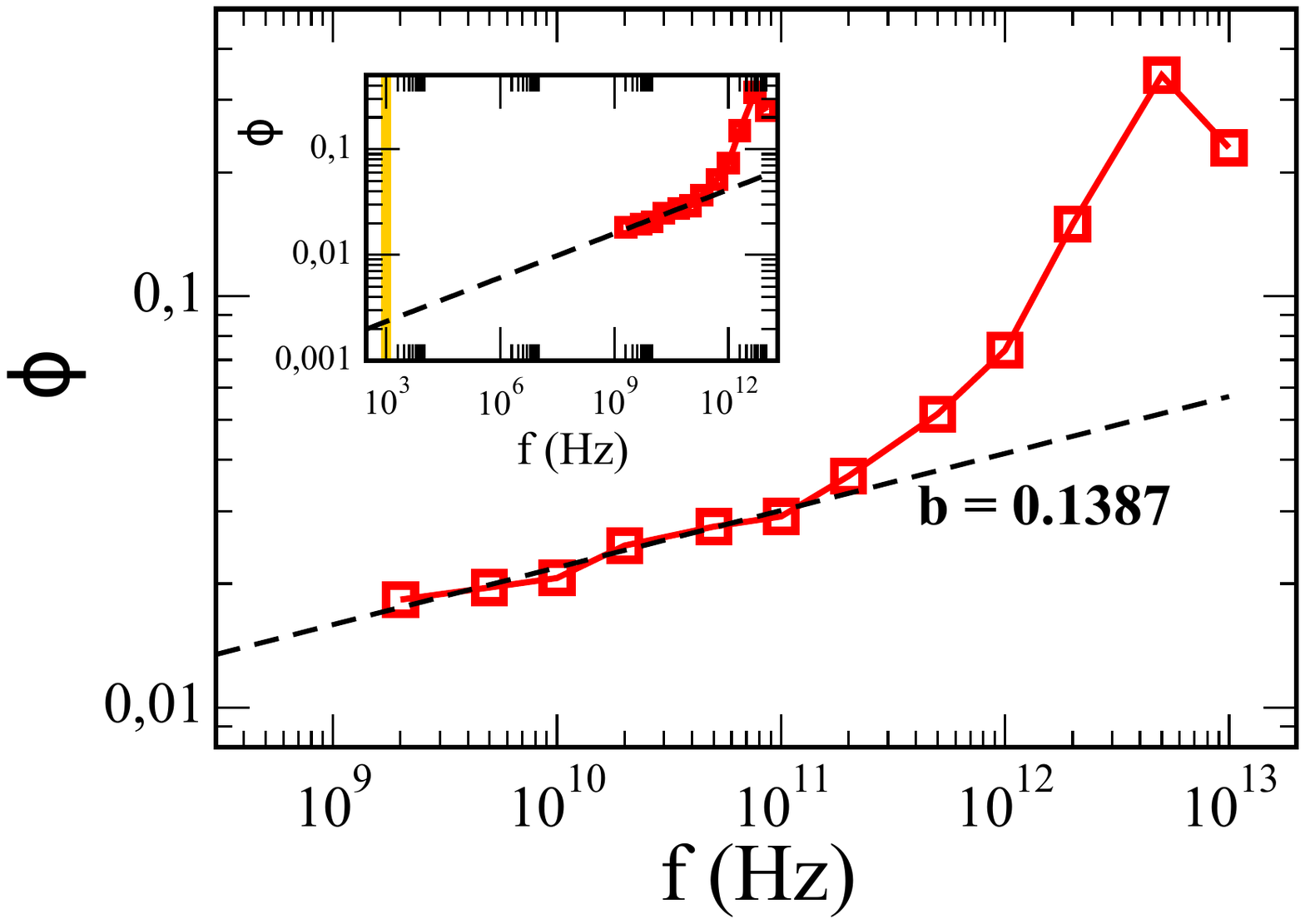}
\caption {Frequency dependence of the coating loss angle  $\varphi_c = Q^{-1}$, where $Q$ is the quality factor obtained from Molecular Dynamics simulations; the dashed line is the best fit curve with a power-law equation $\varphi_c \propto f^{b}$ in the GHz range. {\it Inset}: low-frequency extrapolation of the power-law regime of $\varphi_c$; the vertical line marks the value   $f=1\,\mbox{KHz}$.}
\label{figSimQ}
\end{figure}

First, we calculate the elastic properties of SiC samples. We obtain $E=274\pm1\,\mbox{GPa}$ and $\nu=0.21\pm0.01$ for the Young modulus and the Poisson ratio, respectively, in agreement with values measured on the IBS samples (see Section \ref{SECT_Mech_Prop}). In Figure \ref{figSimQ} we show the frequency dependence of the loss angle $\varphi_c$, computed as the inverse quality factor $Q^{-1}$ in the range accessible to Molecular Dynamics simulations. Unfortunately, this range is quite far from the one studied experimentally in Section \ref{SECT_Mech_Prop}. However, it can be noted that in the GHz range  a well defined power-law regime is found, i.e.,  $\varphi_c = Q^{-1}\propto f^{b}$ with $b=0.14$. Remarkably, the power-law behavior, and in particular the value of the exponent $b$, are in reasonable agreement with the experimental results obtained on the MS samples presented in Section \ref{SECT_Mech_Prop} ($b=0.18$), as well as on previous simulation results on amorphous tantala \cite{Puosi_PhysRevRes2019,Puosi_ActaMat2020}. Assuming the validity of the power-law frequency dependence down to lower frequencies, we may estimate the value of the loss angle in the acoustic region (see the inset of Figure \ref{figSimQ}). We find $\varphi_c( f = 1\,\mbox{kHz})=Q^{-1}_{\tiny{1\,\mbox{kHz}}}=2.34\times10^{-3}$, which is of the same order of magnitude of our experimental results.

\section{Conclusions}
In this work we produced and studied thin films of a-SiC to evaluate the viability of this material for the realization of coatings with ultra-low optical and mechanical losses for their use in ultra-sensitive optical apparatuses, in particular in GWDs. 

Several samples of SiC thin films were successfully grown in their amorphous form, as demonstrated by GIXRD data, using two different sputtering systems. Samples produced via MS exhibited a tendency to grow poor of Si, with a typical Si/C ratio around 0.74, while IBS samples were nearly stoichiometric. The Si deficiency in the MS system could however be easily cured by enriching the Si content on the target. Apart from this aspect, the two set of samples presented a very similar density and a morphology whose differences can be attributed to the different substrates used.

From the optical point of view, our results show that the obtained films present a certain level of variability in the optical constants, which seems however not to be related to differences in composition or on the production method, either MS or IBS. The fact that ageing or annealing can contribute to lower the values of the optical constants suggests that as-deposited samples are in a sort of non-stabilized state and that the observed variability of the optical constants can be attributed to this initial non-equilibrium condition. 
Anyway, all the analyzed a-SiC films display an absorption level which is at least  five orders of magnitude higher than the one of current coatings used in GWDs, such as tantala-titania or silica \cite{Pinard17,Degallaix19}. This seems to indicate that a-SiC thin films are intrinsically much more absorbing in the infrared range than their crystalline bulk counterpart \cite{hofmeister2009optical}, for reasons that appear to be related to their amorphous structure rather than to the presence of dopants or other extrinsic factors.

The mechanical properties were measured in the range of frequencies around 10 kHz. The MS sample exhibited a power law dependency of the loss angle upon the frequency, similarly to what was reported in $\mathrm{Ta}_2\mathrm{O}_5$ and $\mathrm{Ti}:\mathrm{Ta}_2\mathrm{O}_5$ and with a characteristic exponent $b\approx0.18$ which is in line with previous investigations, but about a factor of 2 larger \cite{Granata20}. 
Quite surprisingly, IBS samples tested with the same method produced a qualitatively different result. 
It is not clear yet whether this discrepancy has to be attributed to a real difference in the thin film structure due to the different deposition method, or to some measurement issues. Further investigations on this aspect are ongoing; in particular,as optical measurements indicate that the as-deposited samples appear to be in an initial non-equilibrium condition, it will be interesting to check the effect of annealing procedures on mechanical losses. 
In any case the overall loss angle for all the studied samples is very close $\varphi \approx 10^{-3} \mathrm{rad}^{-1}$ at 10 kHz, which is several times higher than the corresponding value for currently used low-loss high refractive index coatings \cite{Granata20}. 

Finally, our experimental results were used as an input to develop a protocol to simulate the structure and the dynamics of amorphous SiC using a molecular dynamic approach. 
The procedure to generate the sample follows the same protocol already developed to simulate other low-mechanical loss materials \cite{Puosi_PhysRevRes2019, Puosi_ActaMat2020} and consists in simulating a fast quenching of a SiC volume starting from a high-temperature liquid state. 
By selecting the proper cooling rate we succeeded in obtaining a model which is in large part amorphous and displaying a density very close to the experimentally measured one. 
Furthermore, the model was able to correctly reproduce the power-law dependence of the  mechanical losses on the frequency, as it was experimentally observed in our MS samples, with a coherent estimation of the characteristic exponent and of the order-of-magnitude of the losses in the acoustic frequency region, of interest for gravitational wave science. 

The main conclusion of this work is that amorphous SiC, despite its favorable properties, seems not competitive with respect to other materials used in ultra low optical and mechanical loss experiments. 
However the interest in amorphous SiC remains due to its wide range of applications. We thus believe that the data obtained along this work and in particular the molecular dynamics model could serve as valuable resources for further developments in this sense.

\pagebreak

\bibliography{Bibliography.bib}
\bibliographystyle{unsrt}

\end{document}